# Energy and Effective Mass Dependence of Electron Tunnelling Through Multiple Quantum barriers in Different Heterostructures.

Jatindranath Gain[1], Madhumita Das Sarkar[2] and Sudakshina Kundu[2]
[1]Derozio Memorial College, Rajarhat Road, Calcutta-700136; gainelc@gmail.com
[2] Department of Computer Science & Engineering, West Bengal University of Technology, BF-142 Salt Lake Sector-I, Calcutta – 700064, skundu_2003@yahoo.co.in

**Abstract:** Tunneling of electrons through the barriers in heterostructures has been studied, within unified transfer matrix approach. The effect of barrier width on the transmission coefficient of the electrons has been investigated for different pairs of semi conducting materials that are gaining much importance recently. These pairs include CdS/CdSe, AlGaAs/GaAs and InAs/AlSb. Barrier dimensions have been reduced from 20nm to 5nm to observe the effect of scaling on tunneling properties. Material depended is highlighted for electrons with energy varying from below the barrier height to above it. The electron effective mass inside the barrier and the well are often different. The results show that the coupling effect leads to significant changes on the transmission effect. . The effective-mass dependant transmission coefficient has been plotted with respect to electron energy. The computation is based on the transfer matrix method by using MATLAB.

## 1. Introduction

The quantum mechanical tunneling through multiple quantum barriers is a long-standing and well-known problem. Three methods proposed earlier to calculate the tunneling probabilities and energy splitting: (1). Instonton Method (2) WKb Approximation (3) Numerical Calculation.Instaton method is helpful to understand the physical insight of quantum tunneling but the validity is restricted to the case of large separation between the two potential minima. WKB approximation is widely used in its simple mathematical form, but the result is inaccurate due to its inherent defect in connection formula. Recently WKB approximation has been developed by changing the phase lose at the classical turning points but no above approximation have provide the perfect result to the best of knowledge of Author. Using numerical methods, one can get the solution up to the desired accuracy, but a considerable deal of physical insight is lost in this process. In this paper, the Author presented the development of models of multiple quantum wells or barriers potential by using analytical Transfer matrix method (TMM), which has been applied to any arbitrary potential wells and barriers successfully. The author applied the above theory to three electronic device models and got satisfactory results

Low dimensional carrier systems in the semiconductor heterostructures are gaining much importance in recent times due to the potential use of their unique properties in applications ranging from optoeltronics to high speed devices [1-4]. In this connection perpendicular transport of the carriers in semiconductor heterostructures has attracted much attention [5-8]. The MQW structures in particular, are becoming very important due to their potential use in the design and fabrication of quantum cascade lasers, resonant photo detectors, resonant tunneling diodes, single electron tunneling transistors [8] etc. Moreover, with the decrease in the dimensions of the CMOS devices the effect of tunneling of carriers becomes very important in order for estimating the various leakage currents flowing through the devices present in the VLSI chips.

In this paper an attempt has been made to study the tunneling of carriers through a quantum barrier and plot the variation of the transport coefficient with respect to carrier energy. The range of energy include the classically forbidden transistions as well.

The transmission/tunnelling coefficient, which is the flux of particles penetrating through the potential barriers to the flux of particles incident on it at the other interface, has been computed by using the Ben Daniel-Duke (BDD) boundary conditions [9] for solving the Schrodinger equations for the electrons (carriers in this case) inside the coupled well regions and the barrier in between. The theory has been based on the transfer matrix method. Tunneling depends significantly on the barrier width. Scaling of structure dimension affects this variation very sharply.

The material pairs of interest include CdS/CdSe, AlGaAs/GaAs and InAs/AlSb. InAs/AlSbs based HEMTs are excellent for application in satellites due to their low operating voltages [3]. Devices based on AlGaAs/GaAs have been in use over quite some time and CdS/CdSe QW structures promise of improved gain performance for light emitting applications [2]

The effective masses of the carriers are different inside the well and in the barrier, which are made of different materials. Further, this effective mass may change with energy as is given in the case of the InAs/AlSb pair [10]. This affects the tunneling behaviour of the electrons because there is a dependence of the transmission coefficient of the carrier effective masses. In this paper the effect of variation in effective mass on transport coefficient has also been studied. The variation of transport coefficient with electron energy has been studied and compared for different barrier widths for each of these material pairs.

## 2. Theory

The quantum mechanical theory of tunneling through a classically forbidden energy state can be extended for other types of classically forbidden transitions. In this paper a generalised theory of quantum tunneling for transition through multiple quantum barriers has been developed and tested for three different pairs of materials. These adjacent lower energy regions, that are separated by a quantum barrier, are coupled and this is the general pattern for many of the heterostructures. The electron wave functions in the lower energy regions and the barrier region are obtained by solving the Schrodinger equations that satisfy appropriate boundary conditions [11,12]. The solutions depend on the effective masses of the carriers in the regions concerned. Hence the tunneling probability will not only be a function of the dimensions of the barrier alone but will be affected by change of materials as well as its energy dependent effective mass.

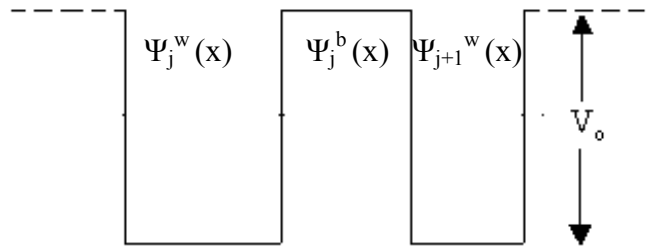

Fig.1.Schematic illustration of a multiple quantum wells (MQW) heterostructure.

The total wave function Φ(r,x) for the electron can be written as

Φ(r,x)=exp(ik$_0$r)Ψ(x)  (1)

Where Ψ(x) satisfies the one dimensional Schrödinger equation with position-dependent electron effective Mass, x represents the growth direction of heterostructure.

The Schrodinger equation according to the effective mass theory for the finite potential barrier and the well regions on either side take on the well-known form

$$\frac{\hbar^2}{2m^*}\frac{d^2\Psi(x)}{dx^2}+(E-V)\Psi(x)=0 \quad (2)$$

with appropriate effective masses m* and potential energies V for the region where the equation is defined. Inside the barrier region the effective mass m*= m$_B$ and potential energy is Vo while those outside the barrier are respectively given by m*= m$_w$ and zero.

Now, we use the transfer matrix for the jth junction and w should generalize the result for N-junctions. This matrix relates the coefficients of the wave function at one end of the junction to those of the other one, so that the wave function can be written as

Ψ$_j^w$(x)=A$_j^w$exp(ik$_w$x)+B$_j^w$exp(-ik$_w$x)  (3)

for the jth well, and

Ψ$_j^b$(x)=A$_j^b$exp(ik$_b$x)+B$_j^b$exp(-ik$_b$x)  (4)

for the jth barrier

Where k$_w$=$\left(\frac{2m_w E}{\hbar^2}\right)^{\frac{1}{2}}$ and k$_b$=$\left[\frac{2m_b(V_0-E)}{\hbar^2}\right]^{\frac{1}{2}}$ with m$_w$ and m$_B$ are the electron effective masses in the regions outside the barrier and barrier region respectively.

By matching the continuity of the wave function Ψ(x) and its appropriate normalised derivative $\frac{1}{m^*}\frac{d\psi(x)}{dx}$ at the boundaries and form 2×2 transfer matrix equations for each interface and we derive a matrix formula that relates the coefficients A$_j$ and B$_j$ with A$_{j+1}$ and B$_{j+1}$.

$$\begin{bmatrix}A_j\\B_j\end{bmatrix}=\begin{bmatrix}M_{11}^{[j]}&M_{12}^{[j]}\\M_{21}^{[j]}&M_{22}^{[j]}\end{bmatrix}\begin{bmatrix}A_{j+1}\\B_{j+1}\end{bmatrix} \quad (5.)$$

By using equation (5) we obtained the coefficients of the wave function at the leftmost slab to those of the right most slab

$$\begin{bmatrix}A^j\\B^j\end{bmatrix}=\frac{1}{2}\begin{bmatrix}1&-ik_w^{-1}\\1&ik_w^{-1}\end{bmatrix}M_j\begin{bmatrix}1&1\\ik_w&-ik_w\end{bmatrix}\begin{bmatrix}A_{j+1}\\B_{j+1}\end{bmatrix} \quad (6)$$

Where $M_j$ is the jth transfer matrix corresponding to the jth junction written as:

$$M_j = M_b(b_j)M_w(a_j)M_b(b_{j+1}) \qquad (7)$$

Where $b_j$ and $a_j$ are the widths of the jth barrier and jth well respectively, $M_b(b_j)$ and $M_w(a_j)$ Correspond to the transfer matrices for the jth barrier and jth well respectively.

Here the total transfer matrix is expressed as the cascading of a series of individual barrier and well.

From equation (6) & (7) it is readily found the transmission amplitude Q is given by

$$Q = \frac{2}{M_{11} + M_{22} + i(k_w M_{12} - k_w^{-1} M_{21})} \qquad (8)$$

Where $M_{ij}$ are the elements of the total transfer matrix.

From the definition of transmission coefficient (T), we obtained the following expression from equation (8)

$$T = |Q|^2 \qquad (9)$$

From equation (5) we obtained the expansion coefficient of the wave function at the left most slabs to those of the right most slab as:

$$\begin{bmatrix} A_1 \\ 0 \end{bmatrix} = M^{[1]}M^{[2]}M^{[3]}\ldots\ldots M^{[j-1]} \begin{bmatrix} A_j \\ 0 \end{bmatrix} \qquad (10)$$

$$= M_{total} \begin{bmatrix} A_j \\ 0 \end{bmatrix} \qquad (11)$$

Which satisfies $M_{21} = 0$     (12)

We obtained the energy eigen values and eigen functions by solving equation (12).

The wave vector $k_w$ outside the barrier is a real quantity for all positive energies E of the electron. The conduction band minimum in the region outside the barrier is taken as the zero of energy. Since the conduction band minimum of the barrier is above that of the region outside hence for certain energies of the electron (E<Vo) the wave vector $k_b$ will be imaginary and for energies above Vo (E>Vo) it will be real. Thus the energy is divided into two regions, (E<Vo) for non-classical transition by tunnelling and (E>Vo) where transition can take place even under classical conditions. The solutions will be different and the methods applied to evaluate the transmission coefficients in the two regions will be different. Since $k_w$ and $k_b$ are both dependent on effective mass and energy, hence for different material pairs the variation of the transmission coefficient for energy values normalised with respect to the barrier height will be different.[13,14]

Here $\beta = m_B/m_w$ is called the mass discontinuity factor. It plays a very important role in determining the transmission coefficient.

For perfect transmission through a barrier sandwiched between two wells we must have

$\beta (k_b/k_w) - k_w/(\beta k_b)]^2 \sinh^2 k_w L = 0$     (13)

Where L is the length of the barrier. So there will be transmission at energy values $E_n = V_o + (n\pi\hbar/L)^2/2m_B$ where transmission coefficient becomes 1. The barrier becomes transparent at these energy values. This is called Ramsauer-Townse [13] effect in atomic physics. The resonance condition is satisfied at normalised energy values $(E/V_o) = 1+n^2/\beta$, where n is an integer. The resonance values therefore depend upon the mass discontinuity factor.

When the electron energy E is less than the barrier height $V_o$ the situation is somewhat different. The above derivations assume that the mass discontinuity factor remains unchanged with energy.

However, it has been pointed out [15] that for InAs/AlSb material system the effective mass varies with the energy according to the expression

$$m(E) = m^*[1+(E-E_c)/E_{eff}] \qquad (14).$$

With $E_c$ represting the conduction band minimum and $E_{eff}$ is the effective band gap at the energy E. The transmission coefficient normally has a dependence on energy through the expression of the wave vector $k_b$. The energy variation of the mass discontinuity factor $\beta(E)$ adds another dimension to this variation. It is worthwhile to examine the effect of this variation and how far it affects the tunneling phenomenon.

The transmission coefficient of electrons through a potential barriers is important for studying the leakage current in MOSFETs with dimensions in the nanometer range. It is also a crucial parameter for studying the behaviour of multiple quantum well structures where the barrier is sandwiched between two coupled quantum wells. In the second case the equations are modified to include the well dimensions also. When both the well and barrier regions are in the nanometer range we expect further quantisation of the energy levels. This is being considered in a further study of the multiple quantum well structure.

## 3. Results and Discussions

In this section we present our results obtained numerically by using MATLAB programming for the transmission coefficient across quantum wire containing multi-barrier heterostructure. The pairs of materials chosen are CdS/CdSe, AlGaAs/GaAs and InAs/AlSb. The parameters used in the computation are given in table 1. The effective mass of electrons in $Al_xGa_{1-x}As$ depends on the mole fraction of x, where x represents the concentration of Al.

In case of $Al_xGa_{1-x}As$ the effective mass is given by $m_{AlGaAs} = (0.063+0.083x)$ and the energy band gap is $E_{AlGaAs}=(1.9+0.125x+0.143x^2)$.

For each pair three values of the barrier width L are taken; 5nm, 10nm and 20nm. In all cases the transmission coefficient increases with diminishing dimension, as expected. This increase is the slowest for InAs/AlSb, which has the highest value of $\beta$. The effective mass variation for AlSb/InAs pair is included in the calculations. It is found that the value of the mass discontinuity factor does not change very sharply with energy for AlSb/InAs. However this slow variation of mass discontinuity factor $\beta$ appears to slow down the increase of transmission coefficient, especially at very low barrier width (5nm).

**Table-1:**

| Parameter | CdS/CdSe | Al$_x$Ga$_{1-x}$As/GaAs (x= 0.47) | AlSb/InAs |
|---|---|---|---|
| Conduction Band-gap $\Delta E_c$ | Eg$_{CdS}$: 2.36eV; Eg$_{CdSe}$: 1.69eV; $\Delta E_c$= 67% of (Eg$_{CdS}$ – Eg$_{CdSe}$) = 0.45 eV | Eg$_{AlGaAs}$: 1.99eV; Eg$_{GaAs}$: 1.42eV $\Delta E_c$=67% x (Eg$_{AlGaAs}$ – Eg$_{GaAs}$) = 0.38 eV | Eco$_{InAs}$: 0.0 eV; Eco$_{AlSb}$: 2.1eV; $\Delta E_c$ = (Eco$_{AlSb}$ – Eco$_{InAs}$) = 2.1 eV |
| Effective mass (m*) | m*$_{CdS}$: 0.20m$_o$; m*$_{CdSe}$: 0.13m$_o$ | m*$_{GaAs}$: 0.067m$_o$; m*$_{AlGaAs}$:0.106m$_o$ | m*$_{InAS}$: 0.020m$_o$; m*$_{AlSb}$: 0.098m$_o$ |
| Mass discontinuity | $\beta$ = 1.54 | $\beta$ =1.58 | $\beta$ = 4.9 |

Quantum tunneling for all barrier widths is least for AlSb barrier in InAs/AlSb pair and most significant for AlGaAs barrier for the in GaAs/AlGaAs/GaAs. This appears to be quite justified because InAs/AlSb/InAs structure has the highest barrier height and GaAs/AlGaAs/GaAs the lowest.(Figures 2, 3, 4)

As the barrier width decreases the tunneling increases and transmission coefficient value rises with normalised electron energy. The rate of rise is sharpest for GaAs/AlGaAs/GaAs and rather slow for InAs/AlSb/InAs. This is easily explained if the barrier heights are compared. CdS/CdSe/CdS structure lies midway.

For ($E_{nor}$ = E/Vo) <1, the transmission coefficient increases from 0 to 1 in a non-linear fashion. For each pair, the transmission coefficient is lower for wider wells as expected. However, the rise is the slowest for InAs/AlSb/InAs because of the highest value of the mass discontinuity factor.

Beyond the normalised energy ($E_{nor}$ = E/Vo) >1, there is resonance; i.e., there are quantised energy values where transmission reaches peak values sharply. This variation is prominent for all three structures, especially near $E_{nor}$ = 1. As the barrier width decreases the peaks get more separated in energy; this is most prominent in GaAs/AlGaAs/GaAs structure. For AlSb/InAs/AlSb the difference between the maxima and minima of the values of the transmission coefficients remain relatively unchanged with decreasing well width. For CdSe/CdS/CdSe and GaAs/AlGaAs/GaAs the maxima of transmission coefficient gradually increase as $E_{nor}$ increases beyond 1.The difference between the maxima and minima also gradually decreases.

Here the regions of lower band gap are taken to be semi-infinite. Hence the normalised energy values vary continuously. If the width of the lower band gap materials outside the barrier are reduced to the order of nanometers then the energy values inside the quantum wells will be quantised. This effect will be reflected in the nature of variation of the transmission coefficients with normalised energy and is expected to change significantly. This will have a crucial effect on carrier tunneling in multiple quantum well structures and leakage current in field effect devices. Studies of this effect are being explored by the author.

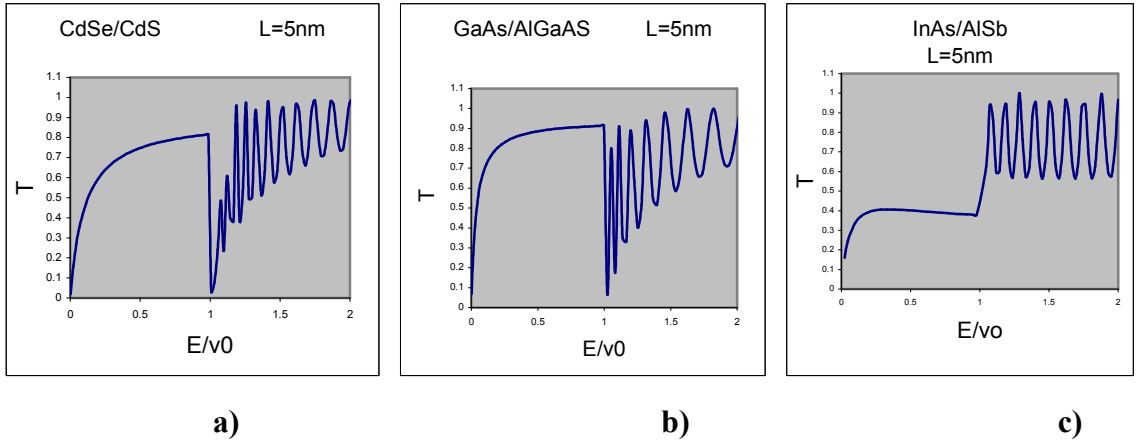

Fig.2: Variation of transmission coefficient of electrons with normalized energy E/Vo for (a) CdSe/Cds (b) GaAs/AlGaAs and (c) InAs/AlSb for well with 5 nm.

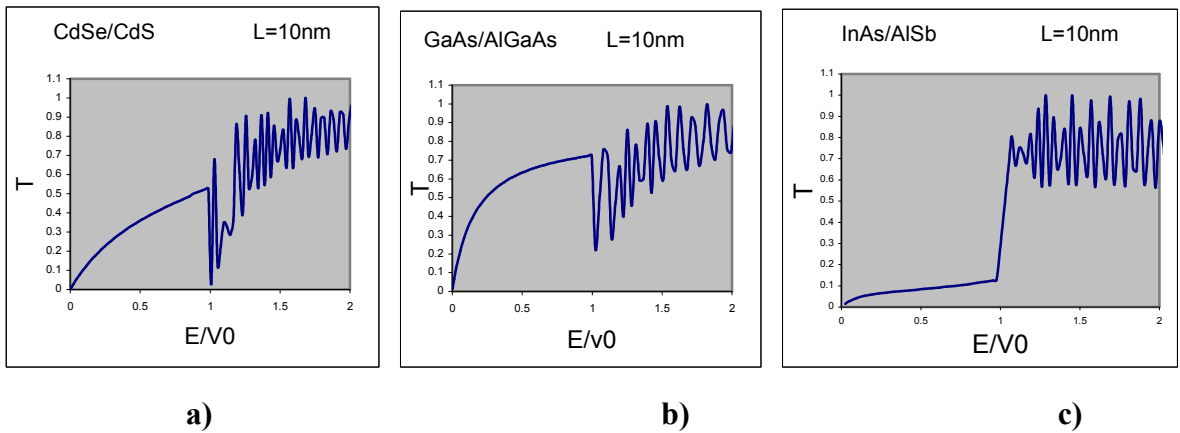

Fig.3: Variation of transmission coefficient of electrons with normalized energy E/Vo for (a) CdSe/Cds (b) GaAs/AlGaAs (c) InAs/AlSb for well with 10 nm.

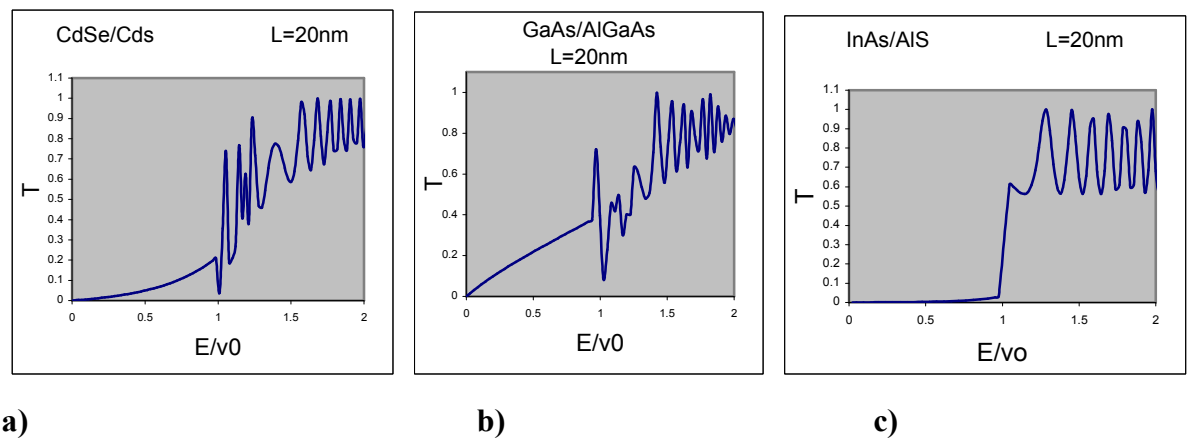

Fig.4: Variation of transmission coefficient of electrons with normalized energy E/Vo for (a) CdSe/Cds (b) GaAs/AlGaAs (c) InAs/AlSb for well with 20 nm

**Acknowledgement**: -This work was financially supported by the University Grant Commission (UGC), India.

# References:

[1] Multilayer Josephson Junction as a Multiple quantum well Structure-A.Hamed Majedi, IEEE Transactions on applied superconductivity, Vol-17, No: 2, June-**2007**

[2] Jeong S. Moon, Rajesh Rajavel, Steven Bui, Danny Wong and David H. Chow, Appl. Phys. Lett, 87, 183110 (2005)

[3] Jianfeng Xu and Min Xiao, Appl. Phys. Lett., 87, 173117 (2005)

[4] B.D.Weaver, J.B.Boos, N.A.Papanicolaou, B.R.Bennet, D.Park and R.Bass, Applied. Phys. Lett. 87, 173501 (2005)

[5] J.Gain, S.Kundu-AOMD-2008

[6] M.Dahan, S.Levi, C.Luccardini, P.Rostaing, B.Riveau and A.Triller, Science, 302, 442 (2003)

[7] V.I.Klimov, A.A.Mikhailovsky, S.Xu, A.Malko, J.A.Hollingsworth, C.A. Leatherdale, H.J.Eisler and M.G.Bawendi, Science, 290, 314 (2000) J.F.Xu, M.Xiao, D.Battaglia and X.Peng, Appl. Phys. Lett., 87, 43107 (2005)

[8] J.Gain, S.Kundu-ICLAN-2006

[9] M.G.Ancona, J.B.Boos, N.Papanicolaou, W.Chang, B.R.Bennet and D.Park, Proc. of Int. Conf. on Simulation of Semiconductor Processes and Devices 2003 (SISPAD 2003) 295, (2003)

[10] K.K.Choi, S.V.Bandara, S.D.Gunapale, W.K.Liu and J.M.Fastenau, Jr. Appl. Phys. 91, 551915 (2002)

[11] D.Barate, R.Tessier, Y.Wang and A.N.Baranov, Appl. Phys. Lett., **87**, 5113 (2005)

[12] Theory of Optical Processes in Semiconductor: Bulk & Microstructure, P.K.Basu, Oxford Science Publication (1997)

[13] Physics of Quantum Well Devices, B.R.Nag, Klewer Academic Publishers.

[14] T.Kuhn and G.Mahler, Physica Scripta, 38, 216 (1988)

[15] Xi.Chen and Chun-Fang Li, Eur. Phys. J. B. 46, 433 (2005)

[16] C.F.Huang, S.D.Chao and D.R.Hang, arXiv: Quant-Ph/0506153 v1, 18 Jun, (2005)

[17] 16.5 μm quantum cascade detector using miniband transport-Fabrizio R.Giorgetta, Petter Krotz and Guido Sonnabend-Applied Physics Letters- June-**2007**